# Using Structurally Well-Defined Norbornyl-Bridged Acene Dimers to Map a Mechanistic Landscape for Correlated Triplet Formation in Singlet Fission

Alexander T. Gilligan, Ethan G. Miller, Tarek Sammakia, Niels H. Damrauer*

Department of Chemistry and Biochemistry, University of Colorado Boulder, Boulder, Colorado, 80309, United States

**Abstract:**

Structurally well-defined TIPS-acetylene substituted tetracene (TIPS-BT1′) and pentacene (TIPS-BP1′) dimers utilizing a [2.2.1] bicyclic norbornyl bridge have been studied – primarily using time-resolved spectroscopic methods – to uncover mechanistic details about primary steps in singlet fission leading to formation of the biexcitonic $^1$TT state as well as decay pathways to the ground state. For TIPS-BP1′ in room temperature toluene, $^1$TT formation is rapid and complete, occurring in 4.4 ps. Decay to the ground state in 100 ns is the primary loss pathway for $^1$TT in this system. For TIPS-BT1′, the $^1$TT is also observed to form rapidly (with a time constant of 5 ps) but in this case it occurs in concert with establishment of an excited state equilibrium (K~1) with the singlet exciton state $S_1$ at an energy of 2.3 eV above the ground state. The equilibrated states survive for 36 ns and are lost to ground state through both radiative and non-radiative pathways via the $S_1$ and non-radiative pathways via the $^1$TT. The rapidity of $^1$TT formation in TIPS-BT1′ is at first glance surprising. However, our analysis suggests that the few-parameter rate constant expression of Marcus theory explains both individual and comparative findings in the set of systems, thus establishing benchmarks for diabatic coupling and reorganization energy needed for efficient $^1$TT formation. Finally, a comparison of TIPS-BT1′ with previous results obtained for a close constitutional isomer (TIPS-BT1) differing in the placement of TIPS-acetylene side groups suggests that the magnitude of exchange interaction in the correlated triplet manifold plays a critical role dictating $^1$TT yield in the tetracenic systems.

**Introduction**

Molecular dimers have emerged as key platforms for the mechanistic exploration of singlet fission (SF),[1,2] and in particular initial photophysics wherein a photoinduced singlet exciton is transformed into a multiexiton state, which is characterized as a singlet-coupled pair of triplets ($^1$TT). Understanding how to control such dynamics is motivated by the premise that SF may serve as a means to down-convert higher energy solar photons into multiple electronic excitations rather than into a single excitation plus waste heat.[3] Dimer and small oligomer systems using acenes,[4-19] but also diimides,[20] and isobenzofurans[21,22] are enabling the interrogation of numerous fundamental issues affecting SF rates and yields, including reaction thermodynamics,[6,16-18,23] state couplings,[8,9,21,24,25] charge transfer intermediates,[10,19-21] the role of entropy,[8] spin dynamics,[15,26] and exciton binding.[8,22,27]

Within the overall body of dimer work in the literature, a leading role has been played by pentacene-based systems[5-7,10-12,14,15,17-19] where the $S_1 \rightarrow {}^1TT$ reaction driving force is significant at -200 to -300 meV and where $^1$TT yields are commonly high, even in the first systems reported.[5-7] A variety of structural motifs have been explored which fall loosely into two groups. In one of these, dimer connectivity occurs via the chromophore ends either using single bonds through the acene 2 position[5,11] or using bicyclic moieties that connect simultaneously through the 2 and 3 positions.[14,17,18] This latter group includes the [2.2.1]-bridge dimer TIPS-BP1′ (see Fig. 1) discussed herein whose synthesis and preliminary photophysics were recently reported by us.[18] In the second group, connectivity occurs at the acene middle, through the 6 position directly[6,10] or via acetylene substituents that then link to a common bridge.[7,12,15,19] While the scope of systems is relatively large and growing, there is not yet consensus about factors controlling important mechanistic details, such as the rate constant for the $S_1 \rightarrow {}^1TT$ forward process. For example, there remain questions about electronic coupling for the photoreaction and whether it is dominated by terms that (a) directly connect the single and double exciton states[28] or (b) demand participation by virtual charge transfer states as is the more common assumption, or (c) entirely system specific. We believe that structurally well-defined dimer systems – including our [2.2.1]-bridge approach and the [2.2.2] and spiropyran approaches of Campos and Sfeir[14] – can play an important clarifying role in the field. By reducing conformational freedom, such systems limit configuration interaction with low energy singlet excimer states.[8,16] As well, they limit uncertainties about state coupling magnitudes and mechanisms that depend on relative chromophore orientation and orientation with respect to bridging moieties. From this vantage point, we would argue that structural definition in dimers provides an opportunity to connect with theory through powerful few-parameter rate expressions such as Marcus theory.[29,30] If this is the case, and if computational tools can be employed to accurately predict physical quantities such as diabatic state couplings, then unifying design principles may have a better chance of emerging.

Although to a lesser extent than the pentacenic systems discussed above, tetraceneic dimers have also been explored and contribute to an overall mechanistic understanding. Early work by Bardeen and coworkers considered phenylene-spaced tetracene dimers.[4] They saw evidence in delayed fluorescence for involvement of the $S_1 \rightarrow {}^1TT$ photoreaction (and its reverse) although they concluded that the $^1TT$ yield was low, of order 3%; notably, that yield can be substantially increase in related systems by introduction of small oligomers such as trimers and tetramers.[31,32] By contrast, Bradforth, Thompson, and coworkers studying highly through-space coupled tetracene dimer systems, saw quantitative conversion of the singlet exciton to a new state that bears both excimer and multiexcitonic ($^1TT$) character.[8] In more rigid and weakly coupled dimers, we initially reported photoluminescence dynamics in room temperature toluene for our [2.2.1]-bridge parent BT1[9] (Fig. 1) and like Bardeen and coworkers concluded that the $^1TT$ yield was low. Our subsequent photophysical studies of a more soluble dimer TIPS-BT1 (Fig. 1) in toluene showed single-exponential singlet-exciton loss concomitant with ground state recovery on the 24 ns time scale and we concluded that the $S_1 \rightarrow {}^1TT$ photoreaction was not operable in that system.[13] We understood this as being a manifestation of point group symmetry properties in the dimer that limits electronic coupling in the photoreaction.[24,33,34] Interestingly, Saito and coworkers recently studied a bent cyclooctatetraene-bridged TIPS-tetracenic dimer with comparable symmetry called FLAP2, and while it has poor photostability compared to its anthracenic and pentancenic analogues, it offers compelling evidence for engaging the $S_1 \rightarrow {}^1TT$ photoreaction on a ps time scale.[17] Those workers note that FLAP2 would have substantially more conformational flexibility about the bridge compared to TIPS-BT1 and suggest that this could lead to the stark dynamical differences between the two dimer systems.

In the work that follows, we explore excited state dynamics for a constitutional isomer of TIPS-BT1 called TIPS-BT1′ (Fig. 1), where the acetylene substitution pattern on each chromophore is moved outwards by a ring relative to the bridge, comparable to what is seen in FLAP2. Transient spectral data offer compelling evidence for the $S_1 \rightarrow {}^1TT$ photoreaction as part of a picosecond timescale equilibration between these states. These data then suggest that the photoreaction energetics are highly sensitive to subtle changes in substitution patterns, for example between TIPS-BT1 and TIPS-BT1′ and lead to marked changes in $^1TT$ yield. Overall, Marcus theory offers a unifying explanation of dynamics in the full set of substituted dimers – TIPS-BT1, TIPS-BT1′, and the pentacenic TIPS-BP1′ – with vibronic coupling derived from symmetry-breaking motions being sufficient to engender fast dynamics.

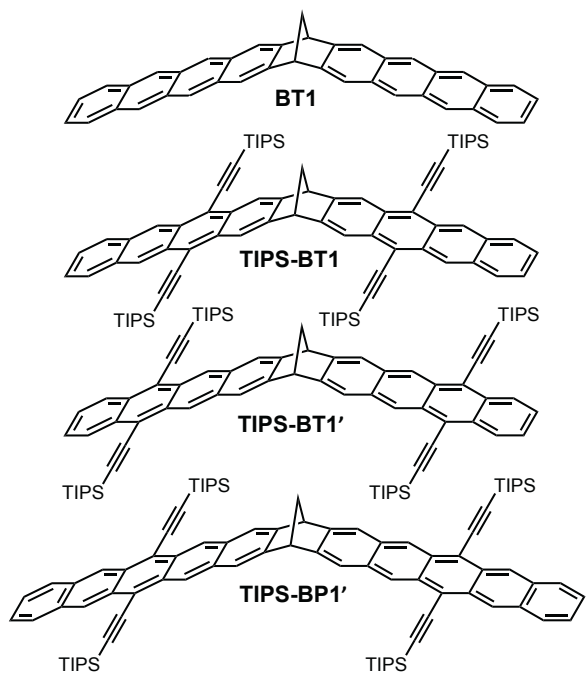

**Figure 1**. Norbornyl-bridged acene dimers discussed in text. BT1 is the conceptual parent.[9] The photophysics of triisopropylsilyl(TIPS)-acetylene substituted bis-tetracene TIPS-BT1 has been explored extensively elsewhere.[9,13] This current work focuses on the substitutional isomer TIPS-BT1′ and the bis-pentacene dimer TIPS-BP1′.[18]

**Results and Discussion**

*Pentacene Dimer*. It is useful to start by characterizing the photoinduced dynamics of TIPS-BP1′ (Fig. 1), a molecule whose reaction driving force is expected to facilitate rapid formation of $^1$TT based on results from a growing number of pentacene-based systems in the literature.[5-7,10-12,14,15,17-19] It is noted that in the communication of our synthetic methodology, we showed preliminary spectral evidence for $^1$TT at 10 ps following photoexcitation.[18] However, that work did not establish time constants or yields. Beginning with ground state absorption, Fig. 2(a) shows a normalized spectrum collected for TIPS-BP1′ in room temperature toluene in a wavelength region that is coincident with our TA measurements described below. To the red is a vibronic progression characteristic of TIPS-Pentacene (TIPS-Pc) moieties, with the 0-0 band peaking at 638 nm. As we have previously described for related systems, the symmetry of this dimer and the fact that the $S_1 \leftarrow S_0$ is acene short-axis polarized, means that only the higher-energy excitonic transition in a Davydov-split pair is bright.[9,13,18] In other words, this system is an H-type aggregate with respect to the $S_1 \leftarrow S_0$ transition of each chromophore arm. To the blue and peaked at 444 nm is a second progression that is also observed in monomer models such as TIPS-Pentacene (TIPS-Pc).[14] Not observed in toluene due to its UV cutoff is the characteristic Davydov splitting associated with coupling the individual-chromophore long axis transitions. As we have shown elsewhere,[18] this feature is readily seen for the molecule in chloroform with intense absorption bands at 308 nm and 333 nm indicating a peak splitting of 0.30 eV. A molar extinction spectrum collected in chloroform is shown in Fig. S1.

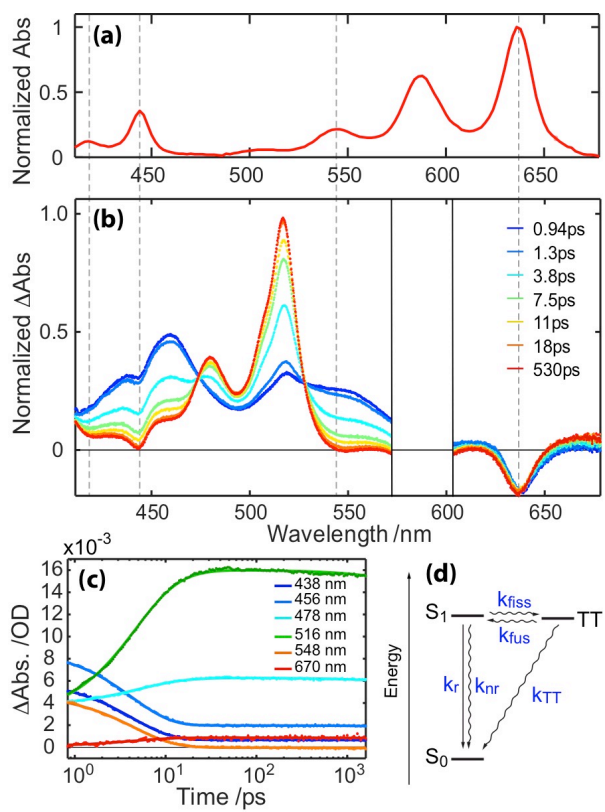

**Figure 2**. (a) Steady-state electronic absorption spectrum of TIPS-BP1′ in toluene at room temperature. (b) Transient absorption spectra of TIPS-BP1′ in room temperature toluene (normalized at Δt = 530 ps). The region surrounding the excitation wavelength of 588 nm is removed due to pump scatter. (c) Selected single wavelength kinetics traces (data points) for TIPS-BP1′ with applied model fits (lines) retrieved from global analysis. (d) Kinetic model of decay pathways of TIPS-BP1' after initial excitation.

Transient absorption (TA) dynamics were collected for TIPS-BP1′ in room temperature toluene following photoexcitation with ~ 50 fs laser pulses at a center wavelength of 588 nm (Fig. 2(b)). The early transient spectrum resembles the lowest energy singlet exciton in a monomer model TIPS-Pc[35] (see Fig. S12) including the excited-state absorption (ESA) at ~ 440 – 470 nm. That spectrum rapidly gives way to a new one that is characterized by the strong ESA at 517 nm along with a vibronic shoulder at 480 nm. These features, which do not further evolve out to the ~ 1 ns limit of this experiment, herald a state with triplet electronic character as seen in a number of SF-active systems involving TIPS-acetylene substituted pentacene chromophores.[5,7] In line with other studies, the speed of the reaction is highly suggestive that the product state is not $T_1$ produced through intersystem crossing, but rather the $^1TT$ produced with spin conserving internal conversion. The data over the time range of 0.5 to 1500 ps are readily fit with a global A→B model with a time constant of 4.4 ps (Fig. S3). It is noted that the strong ESA feature shows a small ~1 nm blueshift over the course of its formation. Although not definitively assigned at this point, it is our expectation that the reactant singlet exciton (state A) is delocalized over both acene arms as was indicated in detailed time-resolved emission studies of TIPS-BT1.[13] In order to estimate the yield of $^1$TT (state B), a sensitization experiment was undertaken to determine the molar extinction of the triplet in TIPS-BP1′, using photoexcited (360 nm) anthracene as a collisional triplet-triplet energy transfer partner (see details in the S.I. and Fig. S8). Here, the assumption is made that that the spectral character of $T_1$ (observed lifetime $\tau_{obs}$ = 55 μs in room temperature toluene; see Fig. S10) is a suitable surrogate for each of the two chromophores in the $^1$TT of TIPS-BP1′. This situation is enabled by the structural rigidity of this dimer, which limits conformational relaxation that might permit significant admixture by other states in the singlet manifold such as excimers.[8,10] With this analysis (see details in the S.I.) we find a yield of 97 ± 11% from the perspective of the $^1$TT or 194 ± 22% from the perspective of triplet excitons (see SI for a description of how error was propagated). These values are in line with those seen in other pentacenic dimer systems.[5,10,12,14]

A longer time resolution TA spectroscopy was used to interrogate the fate of the transient described above that was produced in 4.4 ps. As shown in Fig. S5, the large majority of the signal decays towards baseline with single exponential character and a lifetime of 102 ns. This shortened lifetime for a species that has triplet spectral character (vide supra) is further support for the assignment to $^1$TT.[5,6,15] It is noted that a minor 3.5 % shelf is observed in the time window whose eventual decay to baseline requires 56 μs, thus suggesting assignment to $T_1$. Power-dependent studies did not show a percentage change in the magnitude of the shelf thereby arguing against production of $T_1$ by collision between $^1$TT and ground state species (Fig. S7). It is possible that the shelf manifests as the spin-entangled $^1$TT mixes with the $^5$TT and eventually undergoes decoherence within the dimer into pairs of uncorrelated triplets.[15,26] If this is the case, the shelf would correspond to a dissociated triplet yield of 7% of a possible 200%. A full assignment will require spin-sensitive measurements such as time-resolved EPR.[15,26,36,37]

*Discussion of a Common Model.* As these TIPS-BP1′ studies will help us to understand data in the full series of molecules (Fig. 1), it is useful to present a common framework for analyzing kinetics at this point in the paper. Because of the structural definition of these types of dimers, along with the non-polar solvent environment that precludes significant participation by CT states (vide infra), a relatively simple three-state model can be utilized (Fig. 2(d)).[8,9,20] This includes a singlet exciton state, the TT, and the ground state. The singlet exciton state is coupled directly to the ground state via both radiative and non-radiative pathways ($k_r$ and $k_{nr}$) and it can also be lost due to formation of the TT via $k_{fiss}$ or reformed via the fusion process encompassed in $k_{fus}$. The last rate constant component in this model is the loss pathway linking the TT directly to the ground state, which is referred to as $k_{TT}$. In our understanding of these systems at this time, we assume that TT is primarily the pure singlet $^1$TT produced in the spin-allowed $k_{fiss}$ process, but recognize that this is not an eigenstate of the system[1,38] and that spin mixing with the $^5$TT will begin to occur during the TT lifetime. In a related vein, the model ignores processes leading to the singlet fission product $T_1 + T_1$, which is presumed to occur in conjunction with spin mixing and decoherence, via the $^5$TT. As a common model for each of the dimers explored this is reasonable given that for TIPS-BP1′ the long-time shelf corresponding to this product is relatively small (<3.5 %) and for TIPS-BT1′ it is nearly undetectable.

With this model we can now establish rate constants for the photophysical behavior in TIPS-BP1′. Recalling that the measured $^1$TT yield determined using sensitization experiments is approximately quantitative, a large equilibrium constant K = $k_{fiss}/k_{fus}$ (K ≥ 100) is expected such that the observed exponential decay of 4.4 ps reflects $1/k_{fiss}$ with little contamination (< 1 %) from $k_{fus}$.[39] The large equilibrium constant K also means that the observed 102 ns lifetime of the TA signal has little contamination from $k_r$ and $k_{nr}$ and rather reflects, almost exclusively, $1/k_{TT}$. The values of $k_{fiss}$ and $k_{TT}$ obtained for TIPS-BP1′ are listed in Table 1.

Table 1. Summary of room temperature photophysical properties for dimer species in toluene.

|  | TIPS-BT1 [a] | TIPS-BT1′ [b] | TIPS-BP1′ [b] |
|---|---|---|---|
| $\Phi_{em}$ [c] | 0.72 ± 0.09 | 0.72 ± 0.09 | < 0.01 |
| $\tau_{obs\text{-}fast}$ /ps | 0.85 | 2.5 ± 0.3 | 4.4 ± 0.2 |
| $k_{fiss}$ /s$^{-1}$ | 1.1×10$^{11}$ | (2.0 ± 0.2) ×10$^{11}$ | (2.3 ± 0.1)×10$^{11}$ |
| $k_{fus}$ /s$^{-1}$ | 1.1×10$^{12}$ | (2.0 ± 0.2) ×10$^{11}$ | < (2.2 ± 0.1)×10$^{9}$ |
| $\tau_{obs}$ /ns | 24.3 | 36 ± 3 | 102 ± 3 [d] |
| $k_{TT}$ /s$^{-1}$ | - | - | (9.8 ± 0.3)×10$^{6}$ |
| $\phi(^1TT)$ | ≤ 0.1 | 0.50 ± 0.08 | ≥ 0.97 ± 0.11 |
| $S_1$ /eV | 2.33 | 2.32 | 1.93 |
| $K = k_{fiss}/k_{fus}$ | 0.1 | 1.0 ± 0.1 | 10$^{2}$ - 10$^{5}$ (c.f. 39) |

[a]TIPS-BT1 taken from known value.[13] [b]Reported error is 2σ of three independent measurements. [c]TIPS-BT1′ measured relative to coumarin 540A (coumarin 153) in methanol ($\Phi_{em}$ = 0.45),[40] TIPS-BP1′ measured relative to oxazine 720 (oxazine 170) in methanol ($\Phi_{em}$ = 0.63).[41] [d]Lifetime represents decay of 97 % of initial signal. The remaining signal decays with a lifetime of 56 ± 10 μs.

*Tetracene Dimers*. We next consider the photoinduced dynamics of TIPS-BT1′ whose synthesis follows the same general approach used to prepare the larger acene dimer TIPS-BP1′.[18] As described in the Introduction, we had previously concluded that the close tetracene dimer analog TIPS-BT1 is inactive towards $^1$TT formation as studied in non-polar toluene.[13] As such, our assumption at the outset was that TIPS-BT1′ would also be inactive towards these photophysics due to their structural similarity. This assumption is called into question below.

Steady-state absorption for TIPS-BT1′ in room temperature toluene is shown in Fig. 3 in a spectral region highlighting properties of the lowest energy allowed vibronic transition.

TIPS-BT1′, like TIPS-BP1′ can be characterized as a H-type aggregate with an optically allowed higher energy transition and a dark energetically lower but proximal transition. Also shown in Fig. 3 is the emission spectrum collected for TIPS-BT1′ in the same solvent. The spectrum mirrors the absorption and shows Stokes shifting of 8 nm. From the average of the 0-0 absorption and emission peaks, the value of the optically bright $S_1$ is determined to be 2.32 eV (see Table 1).

As was also the case for TIPS-BP1′, the toluene solvent UV absorption cutoff precludes observation of Davydov coupling between chromophore long-axis transitions. An absorption spectrum collected for TIPS-BT1′ in room temperature chloroform that does show this splitting is presented elsewhere[18] (the molar extinction spectrum is also presented in Fig. S1). In those data, the splitting is 0.47 eV; i.e. a value substantially larger than what is observed for TIPS-BP1′ (0.30 eV, vide supra). It is understood that a significant fraction of the Davydov splitting occurs via Coulomb interaction between individual chromophore transition dipole moments[33,42] and in the case of TIPS-BT1′ those moments have a smaller separation than in TIPS-BP1′.

We next consider a comparison of steady state photophysical data collected for TIPS-BT1′ versus the substitutional isomer TIPS-BT1. Of note, there is very little wavelength shift between these two molecules. The 0-0 transition in TIPS-BT1′ is red-shifted relative to TIPS-BT1 in both absorption and emission data with the bathochromic shift being small (3 nm and 4 nm, respectively). Averaging 0-0 absorption and emission peaks, the optically bright $S_1$ in TIPS-BT1 was determined to be 2.33 eV[13] (see Table 1) or 10 meV higher than what is found in TIPS-BT1′. There are subtle spectral differences between these two molecules that are also worth noting. For TIPS-BT1′, the ratio of 0-0 to 0-1 peak heights in *both* absorption and emission experiments is larger than what is found in TIPS-BT1. For the $S_1$ manifold this is an indication that the two chromophores in TIPS-BT1′ are more weakly interacting than what is seen in TIPS-BT1.[43] Stated a different way, it can be said that in TIPS-BT1′ where the silyl-acetylene groups of the two chromophores are further separated from one-another, the absorptive and emissive transitions are more characteristic of monomer-like line shapes.

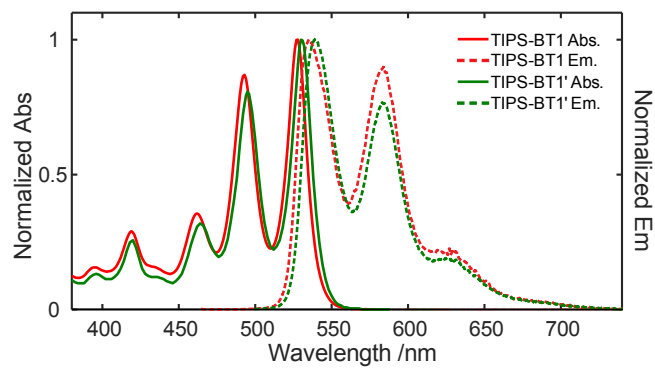

**Figure 3**. Normalized steady-state electronic absorption (solid) and emission (dashed) spectra for TIPS-BT1 (red) & TIPS-BT1' (green) in room temperature toluene.

Time-correlated single photon measurements at 539nm and 584nm, the primary spectral features in Fig. 3, were employed to determine photoluminescence lifetime properties for TIPS-BT1'. The data sets can be modeled using a single exponential decay function with time constant $\tau_{obs}$ = 36 ± 3 ns (see Fig. S2). Notably this observed lifetime is larger than the value recorded for TIPS-BT1 ($\tau_{obs}$ = 24.3 ns[13]) in the same solvent and temperature. Both values are larger than the lifetime collected for the monomer TIPS-Tc ($\tau_{obs}$ = 12.5 ns[13]). We will come back to the lifetime difference between TIPS-BT1' and TIPS-BT1 as it relates to interpretation of an overall decay model for these types of systems.

In our previous communication of synthetic approach to TIPS-BT1' and TIPS-BP1', we reported initial TA spectra collected for these dimers at Δt = 1 ps and 10 ps after photoexcitation over a probe spectral range of 450 nm – 650 nm chosen to interrogate the larger dimer TIPS-BP1'.[18] In that probe range no substantial changes were observed for TIPS-BT1', and this lead us to a preliminarily conclusion that SF dynamics are inactive, in line with our interpretation of photophysics for TIPS-BT1.[13] However, that TA experiment has now been revisited with finer time resolution and using a bluer probe spectrum inspired by the band shape changes observed for TIPS-BP1' in Fig. 2.

TA dynamics for TIPS-BT1' following ~ 50 fs pulse excitation at a center wavelength of 530 nm are shown in Fig. 4(b). Unlike previous measurements for TIPS-BT1 where spectral dynamics were not observed,[13] these new data for TIPS-BT1' show striking evolution within the first ~ 15 ps in spectral regions blue of 450 nm. In particular, rapid loss of intensity is seen for a band in the vicinity of 425 nm, whose line shape is modified by ground state bleach features (see comparison with Fig. 4(a)), but otherwise heralds the $S_1$. Dynamics are seen at other wavelengths as well including significant modification of the magnitude of stimulated emission monitored at ~ 584 nm. Single wavelength kinetic traces extracted from the full spectral data indicate changes in the first 15 ps followed by a lack of further evolution on the 100 ps time scale. The full data set for TIPS-BT1′ inclusive of spectra from $\Delta t$ = 500 fs to 1.5 ns can be modeled using two single exponentially decaying basis functions, one of which has a time constant of 2.5 ps while the second is longer but poorly determined given the time limit of this TA experiment (see modeling discussion in SI and species associated spectra in Fig. S4). We will return to the faster dynamics later and discuss the slower decay first.

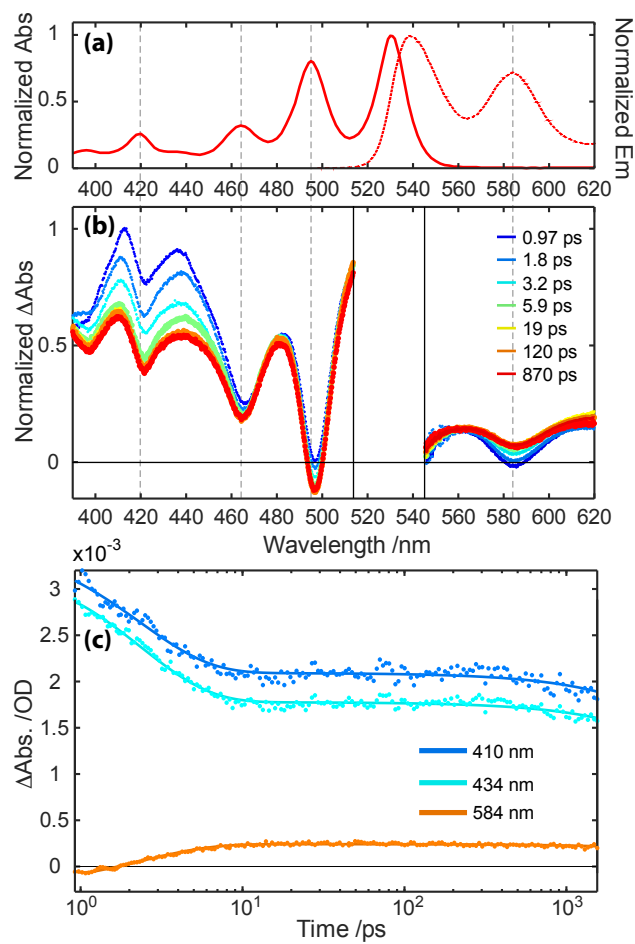

**Figure 4**. (a) Steady-state electronic absorption (solid) and emission (dashed) spectra of dimer TIPS-BT1′ in toluene at room temperature. (b) Transient absorption spectra of TIPS-BT1′ in toluene following ultrafast excitation at 530 nm. The spectral region around the excitation wavelength is removed due to pump scatter. (c) Selected single wavelength kinetics traces (data points) taken from the full-spectrum data with applied model fits (lines) retrieved from global analysis.

To better resolve the slower dynamics, the second TA spectrometer with longer time resolution was again employed. Transient spectral features of TIPS-BT1′ decay to < 1% of baseline and are globally modeled using a single exponential decay with a time constant of 35.8 ns (see Fig. S6). The spectral profile is identical to the second retrieved global fit basis spectrum. This time constant matches the 35 ns lifetime determined from the time-correlated single photon counting studies well (vide supra) and represents ground state recovery.

Returning to the faster 2.5 ps spectral dynamics in Fig. 4, it is noted that the observed changes *cannot* be rationalized by invoking the participation of an intramolecular charge transfer (CT) state formally reducing one chromophore arm of the dimer while oxidizing the other. Whereas population of such a state was previously observed in TIPS-BT1, that measurement required solvation in a polar benzonitrile medium and the results highlighted that excited state equilibrium is established between the CT and a dimer-delocalized singlet exciton at 2.29 eV above ground state.[13] For the same molecule in less polar toluene, where the singlet exciton state is at a similar energy of 2.33 eV, no charge transfer excited state properties are observed.[13] From the perspective of TA spectral changes, the observation of CT for TIPS-BT1 in benzonitrile was very clearly indicated by a transient increase in the magnitude of features tied to the ground-state bleach. This was particularly noticeable at probe wavelengths between ~ 460 nm and 525 nm where singlet exciton ESA features overlap strongly with loss of $S_1 \leftarrow S_0$ absorption: as the singlet exciton ESA is lost in populating the CT, the bleach-related features grow in magnitude with large -ΔA variations. Such changes are absent in TIPS-BT1′ in toluene (Fig. 4(b)) and in fact at a wavelength of 515 nm we observe a small positive change in ΔA as the dynamics unfold.

On the other hand, it is possible to rationalize the transient spectral changes observed for TIPS-BT1′ in Fig. 4(b) if the state being populated has triplet electronic character. Fig. 5(a) presents the Δε spectrum collected for TIPS-BT1′ following triplet sensitization (see SI for experimental details and Fig. S9) which shows two important qualities: first, weak ESA to the blue of 450 nm and second, stronger ESA between 450 nm and 550 nm that is highly modulated with ground-state bleach features leading to the appearance of several positive and negative TA features. The importance of the former is tied to that fact that in in TIPS-BT1′ and in other acetylene-substituted tetracene dimers, the singlet exciton state produced by visible light absorption has a strong ESA in the 400 - 450 nm region. As time evolves and population leaves this state, a weak ESA in the product can accommodate observation of transient loss in ΔA, consistent with what is seen in the first 10 ps (Fig. 4(b)). The importance of the latter ties to our ΔA observations between 460 nm and 550 nm, where changes during the dynamics are actually muted. In TIPS-BT1′, both the nascent singlet exciton and the triplet observed in this region have a strong ESA that is highly modulated by negative peaks associated with ground state bleach (see Δt = 1 ps in Fig. 4(b)). Thus, during interconversion from excited state reactant to product, overall changes in ΔA in this spectral region may in principle be subtle.

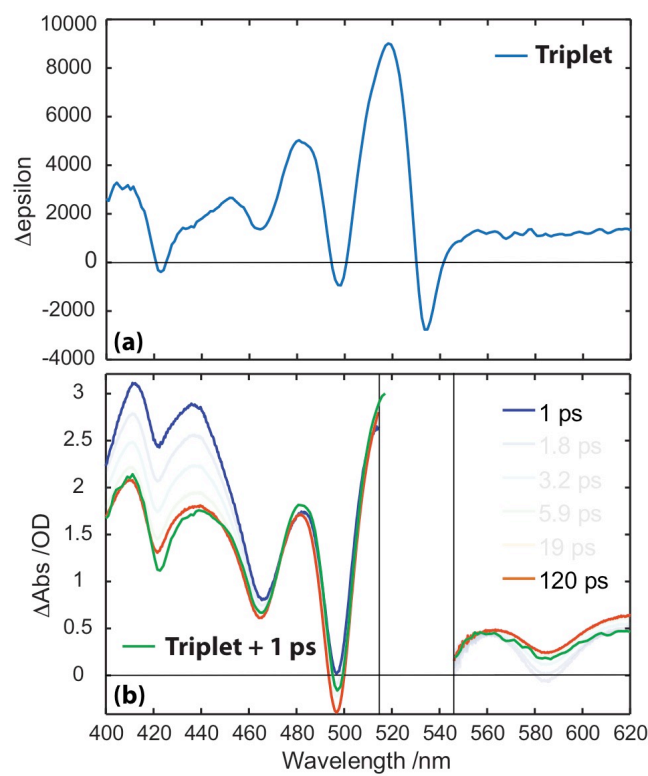

**Figure 5**. (a) Triplet Δε spectrum for TIPS-BT1′ from sensitization experiment in toluene (see SI for sensitization experiment details and Fig. S10) (b) Selected spectral slices for TIPS-BT1′ at 1 ps (blue) and 120 ps (red) along with a reconstructed TA spectrum (green) that is comprised of a superposition between the 1 ps TA spectrum and the sensitized triplet Δε spectrum from (a).

A more quantitative analysis begins by treating later-time spectra – described by the second component retrieved from the global analysis – in terms of two basis functions. The first is a ΔA spectrum collected at early time (Δt = 1 ps) where the dominant contribution is from the singlet exciton whose excited state concentration can be quantified by taking into account the laser power, spot size, and sample absorbance (see SI for details). The second is the triplet Δε spectrum discussed above (Fig. 5(a)). Using a superposition of these two basis functions (50 % singlet exciton and 100 % triplet) we are able to recreate the Δt = 120 ps spectrum with high fidelity as shown in Fig. 5(b). There are two clear implications. The first is that the early dynamics serve to establish an equilibrium between the singlet exciton and a state with triplet character. Given the timescale for the dynamics, that product state cannot be the $T_1$ and rather, is very likely the $^1TT$ where the structural integrity of the bridging norbornyl group enables the two chromophores to essentially preserve their triplet electronic character. This behavior in a tetracene dimer is different than a case where face-to-face interchromophore contact is more intimate leading to significant electronic perturbations.[8] On the other hand it is similar to observations by Saito and coworkers where the chromophores are separated by a bridge derived from cyclooctatetraene.[17,44] The second implication has to do with the basis function percentages needed to reproduce the later-time spectra. The 100% triplet yield needed should be thought of as a 50% TT so the overall population remains conserved in the experiment (50% $S_1$ and 50% $^1TT$). Thus the equilibrium constant established with the 2.5 ps time scale is K = 1.

*Disentangling Dynamics in TIPS-BT1′.* Kinetic modeling using the framework presented in Fig. 2(d) was undertaken for the TIPS-BT1′ data. There are too few independent measurements to uniquely determine each of the rate constants and we choose to draw from information obtained with the other dimers TIPS-BP1′ and TIPS-BT1 in order to gain insight. A starting point is the final decay rate constant $k_{TT}$. In TIPS-BT1′, ground state recovery is strongly influenced by the three rate constants $k_r$, $k_{nr}$, and $k_{TT}$, such that lifetime measurement – even with inclusion of radiative quantum yield information – is insufficient for independent determination of $k_{TT}$. We thus rely on insight from the larger dimer TIPS-BP1′ where $k_{TT}$ was determined to be $1\times10^7$ s$^{-1}$ (Table 1). While useful for modeling purposes, this value is likely an overestimate for TIPS-BT1′. First, $^1TT \rightarrow GS$ is highly exergonic for both dimers (> -1.5 eV) with values that are significantly larger than what would be expected for the reorganization energy of the electronic transformation in each corresponding system. Thus $^1$TT → GS for either TIPS-BP1′ or TIPS-BT1′ is expected to take place in the Marcus inverted region where the reaction should slow as the driving force is increased from TIPS-BP1′ to TIPS-BT1′. Indeed Sanders et al. have observed energy gap law behavior for this decay process in a series of heterodimers.[23] Unfortunately, estimating the extent of the effect in TIPS-BT1′ is further challenged because reorganization energy is also impacted as the acene size is changed. Notably, however, the conclusions reached below are relatively insensitive to the precise value of $k_{TT}$ and we are comfortable setting the value for TIPS-BT1′ at the value measured for TIPS-BP1′.

The next consideration is $k_r$ and $k_{nr}$. Here the dimer system TIPS-BT1 is useful as it is a close structural analog to TIPS-BT1′ but one where $^1$TT formation is minor (K is small) such that the previously reported values of $k_r$ and $k_{nr}$ ($k_r$ = 3.0 × 10$^7$ s$^{-1}$ and $k_{nr}$ = 1.2 × 10$^7$ s$^{-1}$) are the dominant decay paths.[45] Using these $k_r$ and $k_{nr}$ values along with $k_{TT}$ obtained from TIPS-BP1′, the three-state model predicts an observed lifetime for TIPS-BT1′ – that of the S$_1$ ⇆ $^1$TT equilibrium – of 38 ns. This is, in our view, remarkably similar to the kinetic observation of 36 ns (Table 1), thus providing strong support that we understand this TIPS-BT1′ system and that simple three-state model is appropriate.

The final consideration is $k_{fiss}$ and $k_{fus}$. The observed 2.5 ps dynamics in TIPS-BT1′ represents establishment of the S$_1$ ⇆ $^1$TT equilibrium, which then decays in 36 ns. Because of the large separation in these time scales, the rate constant for establishing the equilibrium is simply the sum of $k_{fiss}$ and $k_{fus}$ ($k_{obs-fast}$ = $k_{fiss}$ + $k_{fus}$ = 4.0 × 10$^{11}$ s$^{-1}$). Given K = 1, $^1$TT is both formed and lost with a time constant of 5 ps ($k_{fiss}$ = $k_{fus}$ = 2.0×10$^{11}$ s$^{-1}$). These rate constants were able to accurately reproduce the dynamics of the S$_1$ & $^1$TT populations present in TIPS-BT1′ (see example in Fig. S11). The large $k_{fiss}$ in TIPS-BT1′ was initially surprising to us given aforementioned symmetry issues for this class of dimers.[24] However, diabatic coupling arguments can serve as basis for understanding this rate constant magnitude. In theoretical explorations of vibronic coupling in BT1 – explored because many vibrations break the aforementioned plane of symmetry – we predicted diabatic couplings ($V_{eff}$) between a singlet exciton state and the $^1$TT of order 5.5 meV.[24] Such a quantity is not insignificant inasmuch as it approximately matches what is predicted[24] for tetracene dimer pairs (7.3 meV) germane to the crystal environment where singlet fission is known to take place on the picosecond time scale and be quantitative.[38] Although we have not calculated a comparable $V_{eff}$ value for a TIPS-BT1′ model, we apply the 5.5 meV from the structurally similar BT1 to make rate constant estimates. Using this $V_{eff}$ as the state coupling in Marcus theory, along with a reaction driving force $\Delta G = 0$ meV that is appropriate for a system where K=1, one matches the $k_{fiss} = 2\times10^{11}$ s$^{-1}$ of TIPS-BT1′ when the reorganization energy of the reaction is small, but not unreasonable, at $\lambda = 0.18$ eV.

There is an issue that should be discussed at this point for the sake of completeness. Namely, we have previously argued for BT1 that $\lambda$ for the diabatic $S_0S_1 \rightarrow {}^1TT$ may be larger, of order 0.5 eV.[9] The origin of this prediction is in calculations we made using structures from DFT and TD-DFT with gradients, that predicted a significant intramolecular (inner-sphere) reorganization energy $\lambda_i = 0.43$ eV ($S_1 \rightarrow Q$). Given the current results, this may be an overestimation. We can understand a potential origin of this overestimation in the following way. In our hands, TD-DFT as applied to BT1 and related systems – including use of a toluene solvent continuum model – finds an optimized singlet excited state that is arm-localized. This is true not only for BT1, but also when acetylene substituents are added in respective TIPS-BT1 and TIPS-BT1′ models. However, arm localization contradicts spectroscopic findings for TIPS-BT1[13] where it is apparent that the singlet exciton state for the molecule in toluene is dimer delocalized. We surmise that $\lambda_i$ would be smaller for a dimer-delocalized exciton compared to the arm-localized state found using TD-DFT and additional theory is needed to explore this point. If a lower value of $\lambda$ is operative as is now expected, then we also need to rationalize biexponential photoluminescence behavior observed for BT1.[9] One reasonable explanation, given the poor solubility of BT1 that precluded exploration with TA in the first place, is that aggregation effects contribute to multiexponential decay behavior.

*Comparing TIPS-BT1′ with TIPS-BP1′.* We were initially rather surprised by the overall finding that $k_{fiss}$ for TIPS-BT1′ ($2.0\times10^{11}$ s$^{-1}$) is similar to that of the larger and more exoergic TIPS-BP1′ ($2.3\times10^{11}$ s$^{-1}$; vide supra). As noted earlier for TIPS-BP1′, the $S_1 \rightarrow {}^1TT$ reaction driving force is substantial and expected to be in the -0.2 to -0.35 eV range.[7] However, at the same time the reaction reorganization energy is expected to be smaller in TIPS-BP1′ than the $\lambda = 0.18$ eV suggested above for TIPS-BT1′ given the larger and more highly delocalized chromophores of the pentacenic dimer. Thus for TIPS-BP1′, $S_1 \rightarrow {}^1TT$ conversion is likely to take place in the Marcus inverted region in contrast to the analogous reaction for TIPS-BT1′ and this should contribute to reaction slowing, contrary to our initial assumption. Additionally, whereas the vibronic coupling theory mentioned above predicted diabatic coupling values of order $V_{eff}$ = 5.5 meV for BT1, there is reason to expect it would be smaller in pentacene-based systems where exciton location from the perspective of the individual chromophores of the dimer is moved further away from the bridge linking the two. Qualitatively in support of this, we note our previous observation (vide supra) that Davydov splitting manifest in the UV is smaller for TIPS-BP1′ (0.30 eV) than it is for TIPS-BT1′ (0.47 eV). Factoring each of these things for TIPS-BP1′ – inverted region reactivity and smaller $V_{eff}$ compared to TIPS-BT1′ – it is straightforward to come up with reasonable conditions that give $k_{fiss}$ = 2.3×10$^{11}$ s$^{-1}$.[46] However, given that each Marcus theory parameter is expected to change on going from TIPS-BT1′ to TIPS-BP1′, it is difficult to make specific predictions without further constraints that may come from theory and experiment. Nonetheless, we can emphasize at this point that Marcus theory readily describes the set of behaviors seen in these types of dimer systems.

*Revisiting TIPS-BT1*. As discussed in the Introduction, our published interpretation of TIPS-BT1 photophysics in toluene was that it did not engage in $^1$TT formation and only decayed to ground state via $k_r$ and $k_{nr}$.[13] This was based primarily the lack of spectral evolution in the TA region (~ 420 nm) where there is a strong ESA attributed to the singlet exciton. In that published work, however, we did note a subtle (< 10%) exponential decay of the singlet exciton feature in single-wavelength data ($\lambda_{probe}$ = 429 nm) that was fit with an 850 fs time constant. While the chance of $^1$TT involvement was discussed, it was ultimately dismissed given the stark timescale difference to our BT1 data,[9] and because the absence of spectral evolution argued against it. However, based on the findings herein for TIPS-BT1′, it seems prudent to revisit these conclusions for TIPS-BT1. With the findings for TIPS-BT1′ as a quantitative guide (vide supra), the ~ 10% decay of the $S_1$ magnitude in TIPS-BT1 at $\lambda_{probe}$ = 429 nm is consistent with establishment of a $S_1 \leftrightarrows {}^1$TT equilibrium, but one where the equilibrium constant is small at K ~ 0.1. Using this value in the framework of the three-state kinetic model (Fig. 2(d)), $S_1$ would decay in 850 fs (~10% of signal) as observed if $k_{fiss}$ = 1.07×10$^{11}$ s$^{-1}$ (9.3 ps). This corresponds to an expected slowing relative to TIPS-BT1′ ($k_{fiss}$ = 2.0×10$^{11}$ s$^{-1}$; 5 ps), consistent with the more endergonic driving force of 59 meV (to accommodate K= 0.1). Again Marcus theory is adequate for understanding these results. For example, if $\lambda$ and $V_{eff}$ are respectively held fixed at the previously discussed values of 0.18 eV and 5.5 meV, the time scale for $^1$TT formation in TIPS-BT1 is predicted to be 18 ps; i.e., of the right order of magnitude compared with the 9.3 ps time constant discussed above. Full agreement is achieved if $V_{eff}$ is increased to 7.7 meV. An increase in $V_{eff}$ for TIPS-BT1 relative to TIPS-BT1′ appears to us reasonable, given that the position of the TIPS-acetylene groups influences where the exciton resides, from the perspective of each chromophore relative to the bridge. Qualitative support for a coupling increase is the stronger excitonic interaction observed in TIPS-BT1 compared to TIPS-BT1′ based on vibronic features in the $S_1 \leftarrow S_0$ manifold (vide supra; Fig. 3).

*TIPS-BT1 versus TIPS-BT1′.* As a final point of discussion, we consider how the subtle structural side-group changes that have been implemented manifest in the equilibrium shift from TIPS-BT1 (K = 0.1) to TIPS-BT1′ (K = 1), recalling that this corresponds to a 59 meV exoergic shift for the $S_1 \rightarrow {}^1$TT photoreaction between these two dimers. Some of this could come from state energetics based on observations already discussed. As shown in Fig. 3 and Table 1, the $S_1$ in TIPS-BT1 is slightly higher in energy compared to TIPS-BT1′, by 10 meV. One potential origin of this has to do with electronic perturbations to the acene chromophores that arise from linear attachment to the bicyclic alkyl bridge. In the consideration of monomer models, we have previously shown that the electron-rich bridge serves to modestly destabilize $S_1$ and $T_1$ states relative to pure tetracene, presumably due to electron donating properties of the bridge and their preferential impact on the acene LUMO.[33] In the context of the current dimers, it is reasonable to expect that the position of the TIPS-acetylene substituents will impact the $S_1$ energy, and that this state will be higher for TIPS-BT1 because the acetylene substituents – which participate in determining the average position of the exciton – are closer to the destabilizing bridge. At first glance, the higher $S_1$ might appear to suggest that K would be larger in TIPS-BT1. Importantly however, the same argument applies to the $T_1$ states; i.e., more destabilization in TIPS-BT1 compared to TIPS-BT1′. Assuming the energy perturbation in the triplet manifold is similar to that of the $S_1$,[33] the $S_1 \rightarrow {}^1$TT photoreaction is expected to be more uphill for TIPS-BT1 compared to TIPS-BT1′, given that the energy of the $^1$TT is approximately twice the energy of the $T_1$. However, the extent should be small – of order 10 meV – and while it can contribute, it does not appear significant enough to explain the equilibrium shift observations in total.[47]

A final source of energy perturbation that intrigues us has to do with the biexcitonic TT manifold. As discussed recently by Greenham, Behrends, and coworkers in their electron spin resonance studies of singlet fission in TIPS-tetracene films, triplet interactions in biexciton states are dominated, not by dipolar coupling, but by exchange interactions.[37] The perturbation to the energies of the different state multiplicities that emerge – including the $^1$TT, $^3$TT, and $^5$TT – depends on the extent to which relevant orbitals in the individual chromophore triplets share common space. Unlike dipolar coupling, exchange interactions can account for significant amounts of energy, of order eV, when the extent of common orbital space is extensive as it is in individual acenes; i.e., the reason they are useful for SF problems. Thus even if common orbital space is not large, as one might expect for two acene chromophores juxtaposed relative to one another across a bridge, we suspect that it is not unreasonable to obtain the 10s of meV contributions needed to shift the $S_1 \leftrightarrows$ $^1$TT equilibrium between the two dimers. This would occur by utilizing a combination of through-space interactions as well as through-bond pathways mediated by the norbonyl-bridge σ and σ* system. Such pathways are known to be effective for coupling π−chromophore systems in both electron and energy transfer problems.[42,48-50] In order for this exchange effect to contribute to the observations in our dimer systems, the sign of the TT exchange interaction needs to be controlled such that $^1$TT is destabilized at the same time that $^5$TT is stabilized. This is the same direction one would expect for Hund's rule, and in the case of weakly coupled triplets, this is the direction expected for ferromagnetically coupled electrons across the dimer.[1] In TIPS-BT1′ where the acetylene substituents draw the two triplet excitons further away from one another, exchange interactions would decrease, leading to smaller energy splitting between $^1$TT and $^5$TT and less energetic cost to populating the $^1$TT from the $S_1$ as we have observed. On the other hand, in TIPS-BT1 where the position of the acetylene substituents favors stronger exchange interactions in the TT manifold, the $^1$TT would be pushed to higher energy thus decreasing its relative population within $S_1 \leftrightarrows$ $^1$TT equilibrium, again consistent with our observations. High level electronic structure theory is now needed to confirm the sign of the exchange interaction and to determine the magnitude of the effect in these systems.

**Conclusion:**

In these studies we have focused on two structurally well-defined acene dimers for exploration of excited state dynamics tied to singlet fission. Our emphasis has been on understanding time scales for formation of the multiexcitonic $^1$TT state as well as its loss to the ground state either directly or via pathways involving re-formation and decay of the singlet exciton state. The first dimer system – TIPS-BP1′ – is pentacenic in nature such that $^1$TT formation is exoergic and seen to be efficient with ~ unit quantum yield. The second of these systems – TIPS-BT1′ – is tetracenic and is a close constitutional isomer of a dimer recently studied by our group called TIPS-BT1. The two differ only in the placement of solubilizing TIPS-acetylene side groups. They are energetically quite similar, as borne out using static absorption and emission spectroscopies, and yet they exhibit markedly different evolution of transient absorption features including strong evidence in TIPS-BT1′ for the rapid emergence of significant $^1$TT population.

There are several notable individual findings that are summarized below. However, we first emphasize the general conclusion that in this class of pentacenic and tetracenic dimer systems, where structural definition is by design, we have achieved a unifying understanding of dynamics in terms of the few-parameter rate constant expression of Marcus theory. This allows us to assess appropriate magnitudes for diabatic coupling, reorganization energy $\lambda$, and driving force that enables efficient $^1$TT formation in these and related systems. The overall mechanistic understanding means that these systems can provide benchmarks upon which subsequent variations that alter structure, energetics, and symmetry can be judged.

The first notable specific finding concerns TIPS-BT1′ where we observe rapid formation of the $^1$TT ($\tau_{fiss}$ = 5 ps) in concert with establishment of an excited state equilibrium of equal proportions (K ~ 1) with the singlet exciton state $S_1$ that resides 2.3 eV above the ground state. The established equilibrium means that the $^1$TT resides at a highly similar energy. This speed is initially surprising given the absence of reaction driving force and given the unfavorable structural symmetry in this dimer (a long-axis reflection plane) expected to limit diabatic coupling between reactant and product. [24,33] However, we conclude that we have the framework to rationalize this time constant. Theory we previously applied to the parent norbornyl-bridged tetracene dimer BT1, that factors vibronic coupling through symmetry-breaking vibrational motions (normal modes within the $A_2$ and $B_2$ irreducible representations), predicts an effective diabatic coupling $V_{eff}$ of order 5.5 meV.[24] Such an amount, while appearing to be small, can accommodate $\tau_{fiss}$ = 5 ps without a driving force (appropriate because K ~ 1) when the reorganization energy is low, but entirely reasonable, at $\lambda$ = 0.18 eV. Subsequent theory would be useful to refine these numbers but it is becoming clear that only modest diabatic couplings are needed to enable efficient $^1$TT formation in competition to other radiative and non-radiative decay pathways, in large part because of the small reorganization energies associated with highly delocalized acetylene-substituted acene chromophores engaging in SF. A final point is made about TIPS-BT1′ in relation to the lifetime of the $^1$TT that might be relied upon for subsequent generation of states like the $^5$TT or separated triplets. In this tetracenic system the $^1$TT energy is poised to limit the non-radiative decay to ground state (encompassed in the rate constant $k_{TT}$) compared to pentacenic systems like TIPS-BP1′ that exhibit $^1$TT lifetimes of order 100 ns. Unfortunately, excited state equilibrium with the singlet exciton state undermines this potential gain.

The second notable specific finding concerns the observation that $^1$TT formation in the pentacenic TIPS-BP1′ (4.4 ps) is not substantially faster than in TIPS-BT1′ (5.0 ps) despite the significantly larger (exergonic) reaction driving force of 200 – 350 meV (giving the $^1$TT an energy above the ground-state of ~1.58 – 1.73 eV). This can be partially understood now in the context of Marcus theory where the reaction in TIPS-BP1′ should be slowed by placement in the inverted region. However, other effects are also expected to be in play. Namely, we anticipate reductions in both $\lambda$ and $V_{eff}$ for the more π-delocalized and excitonically separated TIPS-BP1′ relative to TIPS-BT1′ to contribute to the observed similarity in $^1$TT formation rate constants. .

The final notable specific finding concerns the comparison between TIPS-BT1′ and the close constitutional isomer TIPS-BT1 and the fact that despite nearly identical singlet exciton energies, these two molecules exhibit markedly different $^1$TT yields. We believe we are observing the effect of exchange interactions between triplets in the multiexcitonic TT manifold where subtle structural changes – i.e., the placement of the TIPS-acetylene substituents in TIPS-BT1′ versus TIPS-BT1 – are controlling its magnitude and where the comparative observation is revealing its sign. The $^1$TT yields in TIPS-BT1′ versus TIPS-BT1 are consistent with a scenario where exchange interactions raise the energy of the $^1$TT relative to higher multiplicities $^3$TT and $^5$TT. In TIPS-BT1′, the relative placement of the acetylene side groups draws the triplet excitons further away from one another thereby lowering the overall energy of the $^1$TT and enabling its substantial participation (K ~ 1) in equilibrium with the $S_1$ singlet exciton state. The mechanistic details revealed in these comparative studies can be used in the design and interpretation of new systems and architectures to exploit the $^1$TT as a gateway to the $^5$TT or separate triplets.

(39) Based on state energetics, a large K is expected in this system. Note that K = 100 at room temperature for a system with a modest $S_1 \rightarrow {}^1TT$ reaction driving force of -0.12 eV. If the driving force were -0.34 eV as estimated in a related system (c.f. 7) the equilibrium constant K would be greater than $5 \times 10^5$.

(44) It is noted that Saito and coworkers report a nearly quantitative triplet yield of 180%. In our view, this is difficult to reconcile with the reported spectra that show a significant ESA band at ~460 nm as well as significant stimulated emission at ~610 nm, thus signifying the presence of a large amount of excitonic singlet. These workers mention that to account for the dissociation of triplets into independent species from the $^1TT$, the calculated excited-state concentration must be doubled. We believe that this step is unnecessary and that it leads to a triplet yield overestimation while significant amounts of singlets remain. We suspect that the triplet yield in Saito's system is likely closer to the 100% that we report here.

(45) As desribed later, we find K ~ 0.1 in TIPS-BT1 and this means that we are underestimating these $k_r$ and $k_{nr}$ values by a small amount in this molecule by assuming K is very small, as per our original interpretation (c.f. 13). We recover the observed decay with the K = 0.1 in place by setting $k_{tot} = k_r + k_{nr} = 4.5 \times 10^7$ rather than $4.2 \times 10^7$. But this translates to a modest difference in lifetime (22.2ns versus 23.8 ns) and the overall conclusion holds.

(46) For example, setting ΔG = -0.34 eV (c.f. 7) while leaving λ = 0.18 eV, one obtains this value of $k_{fiss}$ when $V_{eff}$ is modestly decreased from 5.5 meV to 4.8 meV.

(47) We have also briefly considered an explanation based on S1 energies. In principle, observed differences in excitonic interactions for TIPS-BT1 versus TIPS-BT1' would manifest in larger energy splitting ($\Delta E_{S1}$) between the higher energy optically bright $S_1$ and the lower energy dark $S_1$. Given that the photoreaction of interest will occur primarily from the lower energy dark $S_1$, the more excitonically coupled TIPS-BT1 could be preferentially disadvantaged. However, in order for this effect to meaningfully lower the equilibrium constant of interest, the difference in $\Delta E_{S1}$ for TIPS-BT1 versus TIPS-BT1' (i.e., $\Delta\Delta E_{S1}$) needs to be a substantial percentage of 59 meV. We do not think this is the case for these dimers. When considering Davydov splitting in the $S_3 \leftarrow S_0$ region, TIPS-BT1 exhibits a larger value (0.499 eV)(c.f. 13) compared to TIPS-BT1' (0.472 eV)(see SI) but this represenst a 5% difference out of ~ 0.5 eV of splitting. If we apply this percentage difference to the much smaller Davydov splitting expected for the $S_1 \leftarrow S_0$ transition in the visible (of order 30 meV; c.f. 33), we find only 1.5 meV to work with. This is not enough to substantively impact the $S_1$ to $^1$TT equilibrium.